\documentclass{article}

\usepackage{psfig,epsfig}
\usepackage{amsmath}
\usepackage{amsfonts,amssymb}

\DeclareMathOperator{\Exp}{Exp}

\long\def\comment#1{}

\newcommand{\R}{{\Bbb R}}
\newcommand{\card}[1]{\left|\;#1\;\right|}
\newcommand{\ind}{\mathcal I}
\newcommand{\mult}{\mathfrak{m}}
\newcommand{\eqdef}{\triangleq}

\newcommand{\NP}{\mbox{\rm NP}}

\newcommand{\OPT}{\mbox{\rm OPT}}

\def\cf{\overline{f}}
\def\ccr{\overline{r}}
\newcommand{\IR}{{\rm\hbox{I\kern-.15em R}}}
\newcommand{\IN}{{\rm\hbox{I\kern-.15em N}}}
\newcommand{\IZ}{{\sf\hbox{Z\kern-.40em Z}}}

\newtheorem{theorem}{Theorem}

\def\qed{\ \hfill\vrule width.2cm height.2cm depth0cm\smallskip}
\newenvironment{proof}{\bigskip\noindent{\bf Proof. }}{\qed}

\newcounter{example}

\newenvironment{algotext}[2]
      {\begin{minipage}{1.0\textwidth}
       \raggedright\sf\noindent
       {\bf Input:} #1\\
       {\bf Output:} #2\\
       \rule[3pt]{1.0\textwidth}{0.3pt}%
       \setlength{\leftmargini}{1.5em}%
       \setlength{\leftmarginii}{2.5em}%
       \setlength{\leftmarginiii}{1em}%
       \begin{description}}
      {\end{description}\end{minipage}}

\newcommand{\step}{\item\hskip-0.4em}

\begin{document}

\title{Approximation Algorithms for Minimum PCR Primer Set Selection
with Amplification Length and Uniqueness Constraints\thanks{Research  
supported in part by a Large Grant from the 
University of Connecticut's Research Foundation.}}

\author{K. Konwar \qquad I. M\u{a}ndoiu \qquad
A. Russell \qquad A. Shvartsman\\[3mm]
University of Connecticut\\
Department of Computer Science \& Engineering\\
371 Fairfield Rd., Unit 2155, Storrs, CT 06269-2155, USA\\
E-mail: \{kishori,ion,acr,aas\}@cse.uconn.edu
}

\maketitle

\begin{abstract}
A critical problem in the emerging high-throughput genotyping protocols is 
to minimize the number of polymerase chain reaction (PCR)
primers required to amplify the single nucleotide polymorphism 
loci of interest.
In this paper we study PCR primer set selection 
with amplification length and uniqueness constraints 
from both theoretical and practical perspectives.  
We give a greedy algorithm that achieves a logarithmic 
approximation factor for the problem of minimizing the 
number of primers subject to a given upperbound on the 
length of PCR amplification products.
We also give, using randomized rounding, 
the first non-trivial approximation algorithm 
for a version of the problem that requires unique amplification 
of each amplification target. Empirical results on 
randomly generated testcases as well as testcases  
extracted from the from the National Center for 
Biotechnology Information's genomic databases
show that our algorithms are highly scalable and produce 
better results compared to previous heuristics.
\end{abstract}

\section{Introduction}
\label{intro.sec}

Availability of full genome data combined with rapid advances in 
high-throughput genomic technologies promises to revolutionize 
medical science by enabling large scale genomic analyses  
such as association studies between Single Nucleotide Polymorphisms 
(SNPs) and susceptibility to common diseases. 
Although recent work \cite{Gabriel02} suggests that there are only a few 
hundred thousand ``blocks'' of SNPs that recombine to provide 
most of the genetic variability seen in human populations, 
meaningful SNP association studies will still require 
genotyping many thousands of SNPs in large populations.\footnote{For 
example, fully powered haplotype association studies are estimated to require 
as much as 300,000 to 1,000,000 ``haplotype-tag'' SNPs \cite{Gabriel02}.}
This poses a daunting challenge to current SNP genotyping protocols
(see \cite{Kwok01} for a survey).  
A critical step in these protocols is the cost-effective amplification 
of DNA sequences containing the SNP loci of interest via 
biochemical reactions such as the {\em Polymerase Chain Reaction} (PCR).  
 

PCR cleverly exploits the DNA replication machinery to 
create up to millions of copies of specific DNA fragments 
(amplification targets). 
In its basic form, PCR requires a pair of oligonucleotides
(short single-stranded DNA sequences called {\em primers}) 
for each amplification target. More precisely, the two 
primers must be (perfect or near perfect) reversed Watson-Crick 
complements of the $3'$ ends of the forward and reverse strands in 
the double-stranded amplification target (see Figure \ref{primers.fig}). 

Typically there is significant freedom in selecting the 
exact ends of an amplification target, i.e., in selecting 
PCR primers. Consequently, primer selection can be 
optimized with respect to various criteria affecting 
reaction efficiency, such as primer length, specificity, 
melting temperature, secondary structure, etc. 
Since the efficiency of PCR amplification falls off exponentially 
as the length of the amplification product increases, 
an important practical constraint is that the binding sites for the 
two primers must be within a certain maximum distance of each other
(typically around 1000 bases).

Much of the previous work on PCR primer selection has focused on 
single primer pair optimization with respect to the above biochemical criteria. 
This line of work has resulted in the release of several  
robust software tools for primer pair selection, the best known of which is
the Primer 3 package \cite{RozenS00}.
Another optimization objective studied in the literature is the 
minimization of the number of PCR primers required to carry out a 
given set of independent amplifications. 
Pearson et al. \cite{PearsonRWZ96} were the first to consider 
this objective in their optimal primer cover problem formulation: 
given a set of DNA sequences and an integer $k$,
find the minimum number of $k$-mers that cover all sequences.  
They showed that the primer cover problem is as hard 
to approximate as set cover, and hence unlikely to be approximable within 
a factor better than $(1-o(1))O(\log{n})$, where $n$ is the number of DNA sequences. 
Pearson et al. also proposed an exact branch-and-bound algorithm 
for the primer cover problem and showed that the classical greedy 
set cover algorithm guarantees a theoretically optimum $O(\log{n})$ 
approximation factor.

{\em Multiplex PCR} (MP-PCR) is a variation of PCR in which multiple DNA 
fragments are amplified simultaneously. Like the basic PCR, 
MP-PCR makes use of two oligonucleotide primers to define the 
boundaries of each amplification target. Note, however, that 
MP-PCR amplified targets are available only as a mixture 
and it may not be possible or cost-effective to separate them 
to the purity required, e.g., in microarray spotting.
Fortunately, this is not limiting the applicability of MP-PCR to 
SNP genotyping, since most of the existing allelic discrimination methods
are highly-parallel and thus can be applied directly to 
mixtures of amplified SNP loci \cite{Kwok01}.  Furthermore, 
effectiveness of allelic discrimination methods is largely 
unaffected by the presence of a small number of 
undesired amplification products, which may occur in MP-PCR. 

A promising approach to further increasing MP-PCR efficiency 
is the use of {\em degenerate PCR primers} 
\cite{Kwok94}.\footnote{Another approach 
is to use PCR primers that complement interspersed 
repetitive sequences, such as the human $Alu$ sequence. 
Since the position of the interspersed
repetitive sequences highly constrains the set of SNP loci 
that can be amplified, this approach is generally not applicable 
when a specific set of SNPs is targeted.}
A degenerate primer is essentially a mixture consisting of multiple 
non-degenerate primers sharing a common pattern and can thus be 
used to simultaneously amplify many 
different SNP loci.  For example, letting $N$ to denote 
a position in the primer sequence 
where all 4 nucleotides can appear in equal proportions, 
the degenerate primer $aNgNc$ represents a mixture of 16 different 
non-degenerate primers ($aagac, aagcc, aaggc, aagtc, \ldots, atgtc$). 
Remarkably, degenerate primer cost is nearly identical to that of non-degenerate 
primers, since the synthesis requires the same number of steps
(the only difference is that one must add multiple  
nucleotides in some of the synthesis steps). However, since not all 
non-degenerate primers present in the degenerate primer mixture 
are useful, it is important to use only degenerate primers with 
bounded degeneracy.
Linhart and Shamir \cite{LinhartS02} proved the NP-hardness of several 
formulations for the degenerate primer design problem, including 
a formulation which asks for a degenerate primer with minimum degeneracy 
that covers a given set of input strings. 
Souvenir et al. \cite{SouvenirBSZ03} proposed an 
iterative beam-search heuristic for the related {\em 
multiple degenerate primer design} problem, which seeks 
a minimum number of degenerate primers, each with bounded degeneracy,
covering a given set of DNA sequences.\footnote{The 
iterative beam-search heuristic of \cite{SouvenirBSZ03} 
is also applicable when a threshold is given for the 
{\em total-degeneracy} of the set of primers rather than individual 
degeneracies.}

A common feature of the string covering formulations 
in \cite{PearsonRWZ96,LinhartS02,SouvenirBSZ03} 
is that they decouple the selection of forward and 
reverse primers, and, in particular, cannot 
explicitly enforce bounds on PCR amplification length.  
Such bounds can be enforced only by conservatively defining the allowable 
primer binding regions (i.e., the DNA segments to be covered). 
For example, in order to guarantee a distance of $L$ between 
the forward and reverse primer binding sites around a SNP, 
one may confine the search to primers binding within $L/2$ nucleotides 
of the SNP locus. However, since this constraint reduces the number of candidate 
primer pairs by a factor of about 2,\footnote{E.g.,
assuming that all DNA $k$-mers can be used as primers, 
out of the $(L-k+1)(L-k+2)/2$ pairs of forward and reverse $k$-mers 
that can feasibly amplify a SNP locus, 
only $(L-k+1)^2/4$ have both $k$-mers within $L/2$ bases of this locus.} 
adopting this approach can lead to significant sub-optimality 
in the number of primers required to amplify all SNP loci.

Motivated by the requirement of unique PCR amplification in  
synthesis of spotted microarrays, Fernandes and Skiena \cite{FernandesS02} 
introduced an elegant {\em minimum multi-colored subgraph} formulation 
for the primer selection problem. In this formulation, each candidate primer is 
represented as a graph node and every two primers that uniquely 
amplify a desired target (e.g., gene) are connected by an edge
labeled (or ``colored'') by the target.  The goal is to find a minimum subset 
of the nodes inducing edges of all possible colors.
Fernandes and Skiena gave practical greedy and densest-subgraph based 
heuristics for the minimum multi-colored subgraph and showed that the problem 
cannot be approximated within a factor better than $(1-o(1))\ln{n} - o(1)$, 
where $n$ is the number of amplification targets. 
While finding a minimum primer set that amplifies a given set of SNPs 
subject to amplification length constraints can be reduced to the  
minimum multi-colored subgraph problem, no non-trivial approximation 
factor is known for the latter problem once unique amplification 
is no longer required. With unique amplification constraints, 
the trivial algorithm of selecting two arbitrary primers 
for each of the $n$ amplification target gives an approximation 
factor of $\sqrt{n}$.


In this paper we study (degenerate and non-degenerate) PCR 
primer selection problems with amplification length and uniqueness  
constraints from both theoretical and practical perspectives.  
Our contributions are as follows:
\begin{itemize}
\item
We give a new string-pair covering formulation for the minimum 
primer set selection with amplification length constraints problem, 
and show that a clever modification of the classical greedy algorithm for 
set cover achieves a near-optimal approximation factor of $\ln(nL)$, 
where $n$ is the number of amplification targets and $L$ is the upperbound 
on PCR amplification length.  This result is complemented by 
a $O(\ln{n})$ inapproximability result, which implies that the 
approximation factor of the greedy algorithm is optimal up to an additive 
term of $O(\ln{L})$
\item 
We give a randomized rounding algorithm with an approximation factor of 
$O(\sqrt{\mult} \log m)$
for the minimum multi-colored subgraph problem 
of \cite{FernandesS02}, 
where $\mult$ is the maximum size of a color class (i.e., 
the maximum number of edges sharing the same color) 
and $m$ is the number of colors. For the minimum
primer set selection with uniqueness constraints 
$\mult=O(L^2)$ and $m=n$. Hence, our result implies 
an approximation factor of $O(L\log n)$, 
which asymptotically improves over the 
trivial approximation bound of $\sqrt{n}$.  Furthermore, our 
algorithm has the same approximation guarantees for the 
minimum multi-colored subgraph problem without  
uniqueness requirements.
\item 
We give the results of a comprehensive experimental study comparing 
our greedy approximation algorithm with previously published 
primer selection algorithms on randomly 
generated testcases as well as testcases extracted from the 
National Center for Biotechnology Information's genomic databases \cite{NCBI}.
\end{itemize}

The rest of the paper is organized as follows. 
In next section we introduce notations and 
give formal problem definitions.  In Section \ref{greedy.sec} we 
describe and analyze the greedy algorithm for the minimum primer 
set selection with amplification length constraints problem.  
In Section \ref{rounding.sec} we give the  
randomized rounding algorithm for the minimum multi-colored subgraph problem.  
Finally, we present experimental results in Section \ref{results.sec} 
and conclude with some open problems in Section \ref{conclusions.sec}.


\section{Notations and Problem Formulations}
\label{formulations.sec}

Let $\Sigma=\{a,c,g,t\}$ be the DNA alphabet. We denote by 
$\Sigma^*$ the set of strings over $\Sigma$, and by $\lambda$ the empty string.
Overloading notations, we use $|\cdot|$ to denote both the 
length of strings over $\Sigma$ and the size of sets.  For a
string $s$ and an integer $t<|s|$, we denote by $s[1..t]$ the prefix of 
length $t$ of $s$.

Following \cite{SouvenirBSZ03}, we define 
a {\em non-degenerate primer of length $k$} as a string from $\Sigma^k$.  
A {\em degenerate nucleotide}  is a non-empty subset of $\Sigma$. 
A {\em degenerate primer of length $k$}, or simply a {\em primer of length $k$}, 
is a string $d_1d_2\ldots d_k$ of degenerate nucleotides, and can equivalently 
be viewed as the set of non-degenerate primers $x_1x_2\ldots x_k$, $x_i\in d_i$.
The {\em degeneracy} of a degenerate primer $d_1d_2\ldots d_k$
is the number of non-degenerate primers it represents, i.e., 
$\prod_{i=1}^{k} |d_i|$.

\begin{figure}[t]
\centerline{\psfig{figure=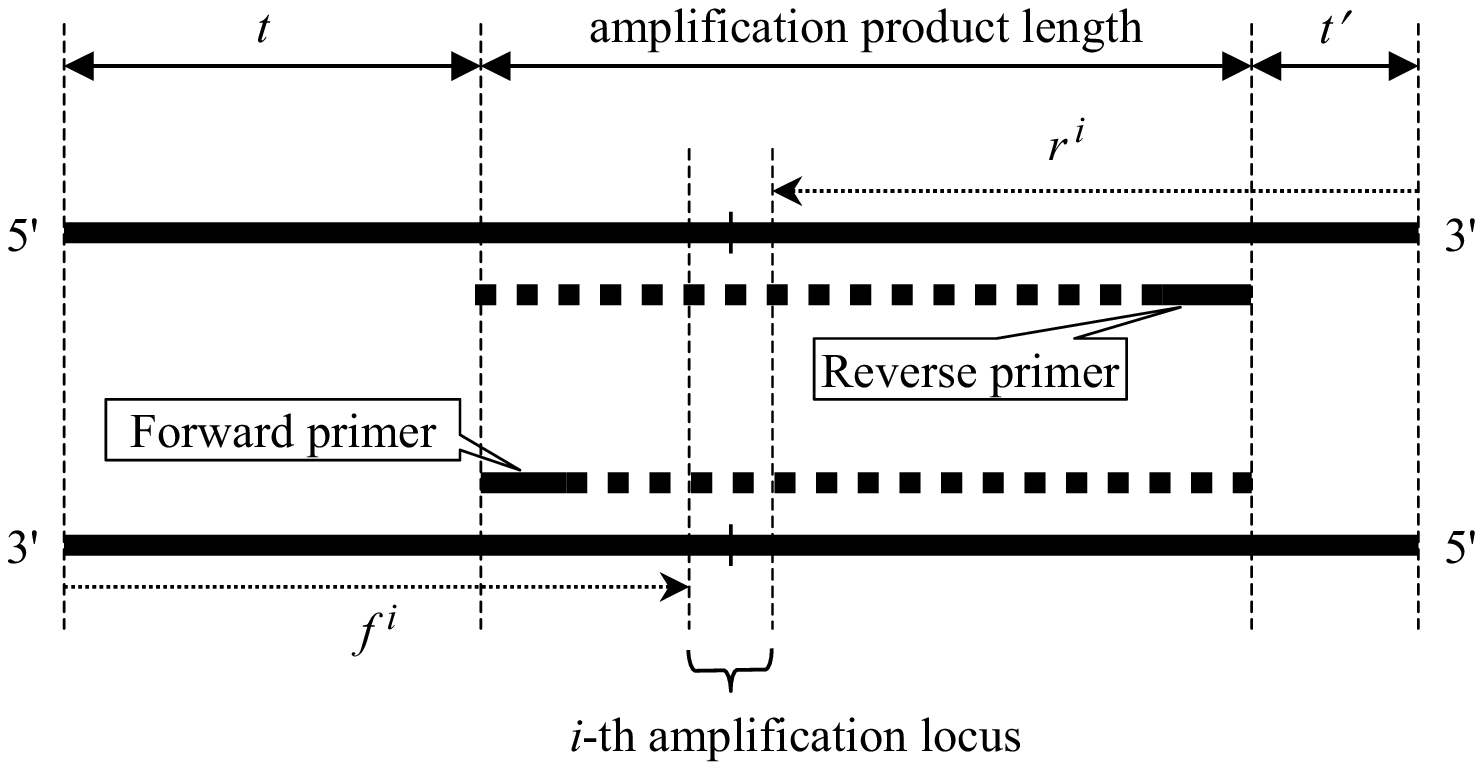,width=3.5in}}
\caption{\label{primers.fig} 
Strings $f^i$ and $r^i$ consist of the $L$ 
DNA bases immediately preceding in $3'-5'$ order
the $i$-th amplification locus along the forward
(respectively reverse) genomic sequence. 
If forward and reverse PCR primers cover $f^i$ and $r^i$
at positions $t$, respectively $t'$, then 
the PCR amplification product length is 
$(2L+x)-(t+t')$, where $x$ is the length of 
the amplification locus ($x=1$ for SNP genotyping).
Thus, amplification product length is at most $L+x$ 
iff $t+t'\ge L$.}
\end{figure}

We denote by $L$ the given threshold on the PCR amplification length, 
and by $f^i$ (respectively $r^i$) the string consisting of  
the $L$ DNA bases immediately preceding in $3'-5'$ order 
the $i$-th amplification locus along the forward 
(respectively reverse) DNA genomic sequence (see Figure 
\ref{primers.fig}). 

We say that degenerate primer $p=d_1d_2\ldots d_k$ 
{\em covers} (or {\em hybridizes at}) 
position $i$ of string $s=s_1s_2\ldots s_m$ 
iff $i$ is the largest index such that 
$s_is_{i+1}\ldots s_{i+k-1}$ is the reversed Watson-Crick complement of one of the 
non-degenerate primers represented by $p$, i.e., iff 
$s_{i+j}$ is the Watson-Crick complement of one of the nucleotides in 
$d_{k-j}$ for every $0\leq j \leq k-1$.\footnote{In practice, 
stable primer hybridization and subsequent 
PCR amplification occur even with a small number of mismatches 
if none of them is too close to the $3'$ end of the primer. 
Our algorithms apply unmodified to hybridization models 
allowing mismatches.}

A set of degenerate primers $P$ is an {\em $L$-restricted primer cover}
for the pairs of sequences $(f^i,r^i)\in \Sigma^L\times \Sigma^L$, $i=1,\ldots,n$, iff 
for every $i=1,\ldots,n$,
there exist primers $p,p'\in P$, not necessarily distinct, 
and integers $t,t'\in \{1,\ldots, L\}$, such that 
\begin{enumerate}
\item $p$ hybridizes at position $t$ of $f^i$;
\item $p'$ hybridizes at position $t'$ of $r^i$; and
\item $t+t'\geq L$
\end{enumerate}
The last constraint ensures that the PCR amplification product length 
is no more than $L+x$, where $x$ is the length of the desired 
amplification target ($x=1$ for SNP genotyping).
We say that a primer cover has the {\em unique amplification property}
if, for each pair $(f^i,r^i)$, there exists exactly one set  
of primers $\{p,p'\}\in P$ satisfying conditions 1-3 above.

The {\em minimum primer set selection problem with amplification length constraints} 
(MPSS-L) is defined as follows:  Given primer length $k$, degeneracy upperbound 
$\delta$, amplification length upperbound $L$, and 
$n$ pairs of sequences $(f^i,r^i)$, $i=1,\ldots,n$, 
find a minimum size $L$-restricted primer cover consisting 
of degenerate primers of length $k$, each 
with degeneracy at most $\delta$.
The {\em minimum primer set selection problem with amplification length 
and uniqueness constraints} 
(MPSS-LU) is defined in the same way except that in this case we seek 
a minimum size $L$-restricted primer cover which has the 
unique amplification property.

\section{The Greedy Algorithm for MPSS-L}
\label{greedy.sec}

\begin{figure}[t]
\fbox{
\begin{algotext}
{Primer length $k$, degeneracy upperbound $\delta$, amplification length upperbound $L$, and
pairs of sequences $(f^i,r^i)\in \Sigma^L\times\Sigma^L$, $i=1,\ldots,n$}
{$L$-restricted primer cover $P$ consisting of degenerate primers of length $k$, 
each with degeneracy at most $\delta$}
\step Function $\Delta(p,i)$:
\step $\Delta \gets 0$
\step If $|\cf^i|+|\ccr^i|\geq L$ return 0
\step If $p$ covers $f^i$ at position $t>|\cf^i|$, 
$\Delta \gets \Delta + (t - |\cf^i|)$
\step If $p$ covers $r^i$ at position $t>|\ccr^i|$, 
$\Delta \gets \Delta + (t - |\ccr^i|)$
\step Return $\min\{ \Delta, L-(|\cf^i|+|\ccr^i|)\}$
\step
\step $P\gets \emptyset$; for every $i=1,\ldots,n$, $\cf^i\gets \ccr^i\gets \lambda$
\step While $\Phi(P):=\sum_{i=1}^n \min\{L, |\cf^i|+|\ccr^i|\} < mL$ do
\begin{description}
\item Find the degenerate primer $p$ maximizing $\Delta\Phi=\sum_{i=1}^n \Delta(p,i)$
\item For every $i=1,\ldots,n$, 
\begin{description}
\item If $p$ covers $f^i$ at position $t>|\cf^i|$ then $\cf^i\gets f^i[1..t]$
\item If $p$ covers $r^i$ at position $t>|\ccr^i|$ then $\ccr^i\gets r^i[1..t]$
\end{description}
\item $P\gets P\cup\{ p \}$
\end{description}
\step Return $P$
\end{algotext}
}
\caption{\label{greedy_algo.fig} The greedy algorithm for MPSS-L}
\end{figure}

MPSS-L can be viewed as a generalization of the partial set cover problem
\cite{Slavik97}.
In the partial set cover problem one must cover with the minimum number 
of sets a given fraction of the total number of elements. In MPSS-L
we can take the elements to be covered to be 
the non-empty prefixes of the $2n$ forward and reverse sequences; 
there are $2nL$ such elements.  
A primer $p$ covers prefix $f^i[1..j]$ ($r^i[1..j]$) if it hybridizes 
to $f^i$ (respectively $r^i$) at position $t\geq j$. 
The objective is to cover at least $L$ (i.e., half) of the elements of 
$\{f^i[1..j], r^i[1..j]~|~1\leq j\leq L\}$ for every $i\in\{1,\ldots,n\}$.

For a set of primers $P$, let $\cf^i$ and $\ccr^i$ denote the longest 
prefix of $f^i$, respectively $r^i$, covered by a primer in $P$. 
Note that $|\cf^i|+|\ccr^i|$ gives the number of elements of 
$\{f^i[1..j], r^i[1..j]~|~1\leq j\leq L\}$ that are covered by $P$.
Let $\Phi(P) := \min\{L, |\cf^i|+|\ccr^i|\}$.  Note that 
$\Phi(\emptyset)=0$, $\Phi(P)=nL$ for every feasible MPSS-L 
solution, and that $\Phi(P)\leq \Phi(P')$ whenever $P\subseteq P'$.
Hence, $\Phi(P)$ can be used as a measure of the progress made 
towards feasibility by a set $P$ of primers.

The greedy algorithm (see Figure \ref{greedy_algo.fig}) 
starts with an empty set of primers and iteratively  
selects primers which give the largest increase in $\Phi$ 
until reaching feasibility. 

\begin{theorem}
\label{greedy_factor.thm}
The greedy algorithm returns an $L$-restricted primer cover of size at most 
$\ln(nL)$ times larger than the optimum.
\end{theorem}

\begin{proof}
Let $\OPT$ denote a minimum size $L$-restricted primer cover, and 
let $p_1,\ldots,p_g$ be the primers selected by the greedy algorithm.
It can be verified that, for every $A$ and $B$, 
$\Phi(A\cup B)\leq \Phi(A) + \sum_{p\in B} [\Phi(A\cup\{p\}) - \Phi(A)]$. 
By using this claim with $A=\{p_1,\ldots,p_{i-1}\}$ and $B=OPT$, 
it follows that in the step when the greedy algorithm selects 
$p_i$, there is a primer in $\OPT\setminus \{p_1,\ldots,p_{i-1}\}$ 
whose selection increases $\Phi$ by at least $(nL - \Phi(P))/|\OPT|$. 
Hence, the selection of $p_i$ must increase $\Phi$ by at least the same amount, 
i.e., reduce the difference between $\Phi(\OPT)$ and $\Phi(P)$ by a factor of 
at least $(1-1/|\OPT|)$. By induction we get that
\begin{equation}
  nL - \Phi(\{p_1,\ldots, p_i\}) \leq nL \left( 1 - \frac{1}{|\OPT|}\right)^{i}
\label{subadd.eq}
\end{equation}
which implies that the number of primers selected by the greedy algorithm is 
at most $\ln(nL)$.
\end{proof}

\noindent
{\bf Remark. }
In \cite{PearsonRWZ96} it is proved that the following primer cover 
problem is as hard to approximate as set cover:
Given integer $k$ and strings $s_1,\ldots,s_n$, 
find a minimum set of $k$-length primers covering all $s_i$'s.
A simple approximation preserving reduction of the 
primer cover problem to MPSS-L shows 
that the MPSS-L problem cannot be approximated 
within a factor better than $(1-o(1))\ln{n}$ unless 
$\NP \subseteq \mbox{TIME}(n^{O(\log \log n)})$.
Hence, the approximation factor in Theorem \ref{greedy_factor.thm}
is tight up to an additive term of $O(\ln{L})$.

\section{Rounding Algorithm for the Minimum Multi-Colored Subgraph Problem}
\label{rounding.sec}

In this section we consider a graph-theoretical generalization of the
MPSS-LU problem. The {\em minimum multi-colored subgraph} problem
\cite{FernandesS02} is defined as follows. Let $G = (V, E)$ be an
undirected graph and $\chi_1, \ldots, \chi_k \subset E$ a family of nonempty ``color
classes'' of edges with the property that $\bigcup_i \chi_i = E$.  Assigning $X
= (\chi_1, \ldots, \chi_k)$, let $\ind(G, X)$ denote the minimum size of a set of
vertices $I$ for which the subgraph induced by these vertices contains
at least one edge of each color. Note that $2 \leq \ind(G,X) \leq 2|X|$ and,
as an edge may belong to several distinct color classes, both of these
extreme values are in fact possible.

The problem of computing $\ind(G,X)$ is \textsc{np}-hard, via, e.g., a
natural reduction from \textsc{set-cover}. We show below that it can
be approximated to within $O(\sqrt{\max_{\chi} \card{\chi}} \log \card{X})$ in
polynomial time.

\begin{theorem}
  $\ind(G,X)$ can be approximated to within $O(\sqrt{\mult} \log
  \card{X})$ in polynomial time, where $\mult = \max_{\chi \in X} \card{\chi}$.
\end{theorem}

\begin{proof}
  We begin with the following integer program formulation of this
  optimization problem
  \begin{align*}
    \min \sum_v x_v,& \;  \text{subject to}\\
    &\forall \chi \in X, \sum_{e \in \chi} y_e \geq 1\enspace,\\
    &\forall v \in V, \forall \chi \in X, \sum_{v \in e \in \chi} y_e \leq x_v\enspace,\\
    &\forall e \in E, y_e \geq 0, \forall v \in V, x_v \geq 0\enspace.
  \end{align*}
  Relaxing this formulation by allowing the variables $x_v$ and $y_e$
  to take values in $[0,1]$ results in a linear program, the optimum
  value for which we denote $\ind_\ell(G, X)$. We begin by scaling the
  linear program to obtain the following new linear program:
  \begin{align*}
    \min \sum_v x_v,& \;  \text{subject to}\\
    &\forall \chi \in X, \sum_{e \in \chi} y_e \geq \sqrt{\mult}\enspace,\\
    &\forall v \in V, \forall \chi \in X, \sum_{v \in e \in \chi} y_e \leq x_v\enspace,\\
    &\forall e \in E, y_e \geq 0, \forall v \in V, x_v \geq 0\enspace.
  \end{align*}
  Let $\ind_{\ell}^s(G,X)$ denote the optimum value for this scaled
  version, and note that $\ind_{\ell}^s(G,X) \leq \sqrt{\mult}\cdot \ind_\ell(G,X)$
  by scaling any solution that achieves the value $\ind_\ell(G,X)$ by the
  factor $\sqrt{\mult}$; let $x^* \in \R^V$ and $y^* \in \R^E$ denote a
  feasible solution to the program above, achieving the optimum value
  $I_\ell^s(G,X)$.

  Based on the solution $(x^*,y^*)$ above, define a family of
  (artificial) independent $\{0,1\}$-valued random variables 
  $$
  \{ Z_{v,e} \mid v \in e, v \in V, e \in E\}
  $$
  where $\Pr[Z_{v,e} = 1] = p_e \eqdef \min(y_e^*,1)$ for each $v \in
  e$. In terms of these variables, define, for each $v \in V$ and each
  $(u,v) = e \in E$, the variables
  $$
  X_v = \bigvee_{v \in e \in E} Z_{v,e} \qquad \text{and} \qquad Y_{e} = Z_{u,e}Z_{v,e}\enspace.
  $$
  Finally, we let the variables $X_u$ determine a random set of
  vertices $S = \{ v \mid X_v = 1\}$. Our goal is to show that, for each
  color class $\chi$, the set $S$ is likely to induce an edge in $\chi$.
  
  \emph{Comment}. Observe that indicator variable for the event that
  the set $S$ induces the edge $e = (u,v)$ is $X_uX_v$ which dominates
  the variable $Y_{(u,v)}$. We focus on this second, less natural, set
  of variables because, unlike the variables $X_uX_v$, the $Y_{(u,v)}$
  are independent.
  
  With this in mind, note that $\Pr[Y_e = 1] = (p_e)^2$ and that for
  each $v$
  \begin{align*}
    \Pr[v \in S] &= \Pr\left[ X_v = 1\right] = \left(1 - \prod_{v \in e} \Pr[Z_{v,e} =
      0]\right) = \left(1 - \prod_{v \in e} (1 - p_e)\right)\\
    &\leq \left(1 - \Bigl(1 - \sum_{v \in e} p_e\Bigr)\right) \leq \sum_{v \in e}
    y_e^* \leq x_v^*\enspace.
  \end{align*}
  Hence, by linearity of expectation
  $$
  \Exp\left[ |S| \right] =\Exp\left[\sum_v X_v\right] \leq \ind_\ell^s(G,X) \leq \sqrt{\mult}\cdot
  \ind_\ell(G,X) \leq \sqrt{\mult}\cdot \ind(G,X)\enspace.
  $$
  We wish to upper bound, for each color class $\chi$, the quantity
  $$
  \Pr [\forall e \in \chi, Y_e = 0] = \Pr\left[S\;\text{induces no edge from}\;\chi\right]
  $$
  with the intention of showing that this selection $S$ of vertices
  is likely to induce many color classes. So, consider now an
  arbitrary color class $\chi$; then 
  $$
  \Exp\left[\sum_{e \in \chi} X_uX_v\right] \geq \Exp\left[\sum_{e \in \chi} Y_e\right] = \sum_{e \in \chi} p_e^2 \geq
  \card{\chi} \cdot \left( \frac{\sqrt{\mult}}{\card{\chi}}\right)^2 \geq 1\enspace,
  $$
  as $\sum_{e \in \chi} p_e \geq \sqrt{\mult}$ and the function $x \mapsto x^2$ is
  convex. Considering that the $Y_e$ are independent, we compute
  \begin{align*}
    \Pr[\chi\;\text{not induced by}\;S] &= \Pr[\forall {(u,v) \in \chi}, X_uX_v = 0]
    \leq \Pr\left[\forall e \in \chi, Y_e = 0\right]\\
    &= \prod_{e \in \chi} (1 - p_e^2) \leq \prod_{e \in \chi} e^{-p_e^2} = e^{-\sum_{e \in \chi}
      p_e^2 }\geq e^{-1}\enspace.
  \end{align*}
  Evidently, selection of $S$ as above ``covers'' any individual class
  $\chi$ with constant probability. So, finally, consider the set of
  vertices obtained by (i.) repeating the above procedure $t = (\log
  \card{X}+2)$ times, resulting in the vertex sets $S_1, \ldots, S_t$ followed
  by (ii.) forming the union $S = \bigcup_i S_i$. Then
  $$
  \Exp[ \card{S}] \leq  \sqrt{\mult} (\log \card{X} +2) \cdot \ind(G,X)
  $$
  so that by Markov's inequality, the probability that $\card{S}$
  exceeds this value by a factor $3$ is no more than $1/3$. In
  addition, the probability that $S$ fails to induce an edge in all of
  the color classes is
  $$
  \Pr[\exists \chi \in X, \text{no edge of $\chi$ induced by}\;S] \leq \card{X} \cdot
  \left(e^{-1}\right)^{\log\card{X}+2} = e^{-2} \leq 1/3\enspace.
  $$
  Hence with constant probability this procedure results in a
  collection of vertices that induces at least one edge of each color
  class and has cardinality no more than $O(\sqrt{\mult} \log
  \card{X}) \ind(G,X)$, as desired.
\end{proof}

We show below that the integrality gap of the LP defining
$\ind_\ell(G,X)$ is $\Omega(\sqrt{\mult})$ in general. This suggests that this
particular LP formulation may have limited value in achieving
approximation results beyond the $\sqrt{\mult}$ threshold.

\begin{theorem}
  For every $s \geq 0$ there is a pair $(G, X)$ for which $\mult = s$ and
  $\ind(G,X) \geq \Omega(\sqrt{\mult}) \ind_\ell(G, X)$.
\end{theorem}

\begin{proof}
  Consider the graph on $n \gg s$ vertices obtained by selecting,
  independently and uniformly at random, $n$ matchings $\chi_1, \ldots, \chi_n$
  each of size $s$ and assigning $E = \bigcup \chi_i$. Observe that the
  feasible solution obtained by setting $x_v = y_e = 1/s$ for all $e$
  and $v$ implies that $\ind_\ell(G,X) \leq n/s$.
  
  On the other hand, we show that with high probability, this random
  selection of matchings results in a graph for which the smallest
  integer solution has objective value at least $\ell \eqdef (n-1) /
  \sqrt{2s}$.  Specifically, let $L \subset V$ be a fixed collection of $\ell$
  vertices and note that the probability that any given edge induced
  by $L$ is included in, e.g., $\chi_1$ is $s / \binom{n}{2}$; hence the
  probability that $L$ induces an edge of each color is no more than
  $$
  \left( \frac{s}{\binom{n}{2}} \binom{\ell}{2} \right)^m \leq \left(\frac{s
      \ell^2}{(n-1)^2}\right)^m \leq \left(\frac{1}{2}\right)^m\enspace.
  $$
  Hence the probability that some set of $\ell$ vertices induces an
  edge of each color is no more than $\binom{n}{\ell} 2^{-m} < 1$ for $m
  \geq n$. Evidently, there exists a family of color classes $X = (\chi_1,
  \ldots, \chi_m)$ for which $\ind(G,X) \geq \Theta(\sqrt{\mult}) \ind_\ell(G,X)$, as desired.
\end{proof}

\section{Experimental Results}
\label{results.sec}

We performed experiments on both randomly generated MPSS-L 
instances and instances extracted from the human genome databases.
Random DNA sequences were generated from the uniform distribution 
induced by assigning equal probabilities for each nucleotide.
The DNA sequences consisted of regions surrounding 100 known
SNPs collected from National Center for Biotechnology Information's 
genomic databases \cite{NCBI}. 

For all experiments we used a bound $L=1000$ on the PCR amplification length. 
In all experiments we considered only non-degenerate primers
($\delta=1$) with length $k$ between 8 and 12. 
These values model the restricted degenerate primer format suggested 
and experimentally validated by Jordan et al. \cite{Jordan02}. 
In this format, 8-12 nucleotides at the $3'$ end of 
each primer are fully specified, followed by a middle sequence of 
up to 6 fully degenerate nucleotides, followed by a fixed 
GC-rich sequence (CTCGAG in \cite{Jordan02}) 
at the $5'$ end.

We compared the following four algorithms:
\begin{itemize}
\item  
The greedy primer cover algorithm of \cite{PearsonRWZ96} (G-FIX).
In this algorithm the candidate primers are collected 
from the reverse and forward sequences within a distance of 
$L/2$ around the SNP.
This ensures that our final  solution is a set of primers that 
meets the product length constraints. The algorithm repeatedly selects 
the candidate primer that covers the maximum number of not yet covered 
forward and reverse sequences. 

\item  

A na\"{\i}ve modification of G-FIX, which we call G-VAR, 
in which the candidate primers are initially collected 
from the reverse and forward sequences within a distance of 
$L$ around the SNP. The algorithm proceeds by greedily 
selecting primers like G-FIX, except that after a first 
primer $p$ covers one of the forward or reverse sequences 
corresponding to a SNP at position $t$, we truncate the 
opposite sequence to a length of $L-t$, thus ensuring 
that the final primer cover is $L$-restricted.

\item  
The greedy approximation algorithm from Figure \ref{greedy_algo.fig}, 
called G-POT since it makes greedy choices based on the 
``potential function'' $\Phi$.

\item  
The iterative beam-search heuristic of Souvenir et al. \cite{SouvenirBSZ03}. 
We used the primer-threshold version of this heuristic, MIPS-PT, with 
degeneracy bound set to 1 and the default beam size of 100.
\end{itemize}

Table  \ref{table_greedy} gives the number of primers selected and
the running time (in CPU seconds) for the three greedy algorithms
and for the iterative beam-search MIPS-PT heuristic of\cite{SouvenirBSZ03}
on instances extracted from the NCBI repository.
G-POT has the best performance on all testcases,
reducing the number of primers by up to 24\%  compared to
G-FIX and up to 30\% compared to G-VAR.  G-VAR performance
is neither dominated nor dominating that of G-FIX.
On the other hand, the much slower MIPS-PT heuristic has
the poorest performance, possibly because is fine-tuned to
perform well with higher degeneracy primers.

\setlength{\tabcolsep}{2pt}

\begin{table}[t]
{\footnotesize 

\begin{tabular}{| c | c|c c | c c| c c | c c |}
\hline
\# & $k$ &
\multicolumn{2}{c|}{G-FIX} &   
\multicolumn{2}{c|}{G-VAR} &
\multicolumn{2}{c|}{MIPS-PT} & 
\multicolumn{2}{c|}{G-POT}       
\\
SNPs & &
\#Primers & CPU sec. &
\#Primers & CPU sec. &
\#Primers & CPU sec. &
\#Primers & CPU sec.\\
\hline                
\hline
 50  &   8 & 13 & 0.13 &  15  & 0.30  & 21  &   48  & 10   & 0.32
\\
 50  &  10 & 23 & 0.22 &  24  & 0.36  & 30  &  150  & 18   & 0.33
\\
 50  &  12 & 31 & 0.14 &  32  & 0.30  & 41  &  246  & 29   & 0.28
\\
100  &   8 & 17 & 0.49 &  20  & 0.89  & 32 &  226  & 14   & 0.58
\\
100  &  10 & 37 & 0.37 &  37  & 0.72  & 50  & 844  & 31   & 0.75
\\
100  &  12 & 53 & 0.59 &  48  & 0.84  & 75  & 2601 & 42   & 0.61
\\
\hline
\end{tabular}
\caption{\label{table_greedy} Results 
on instances extracted from NCBI repository ($L=1000$).}
}
\end{table}

\begin{figure}[t]
\begin{center}
    \begin{minipage}[b]{0.46\linewidth}
      \centering \psfig{figure=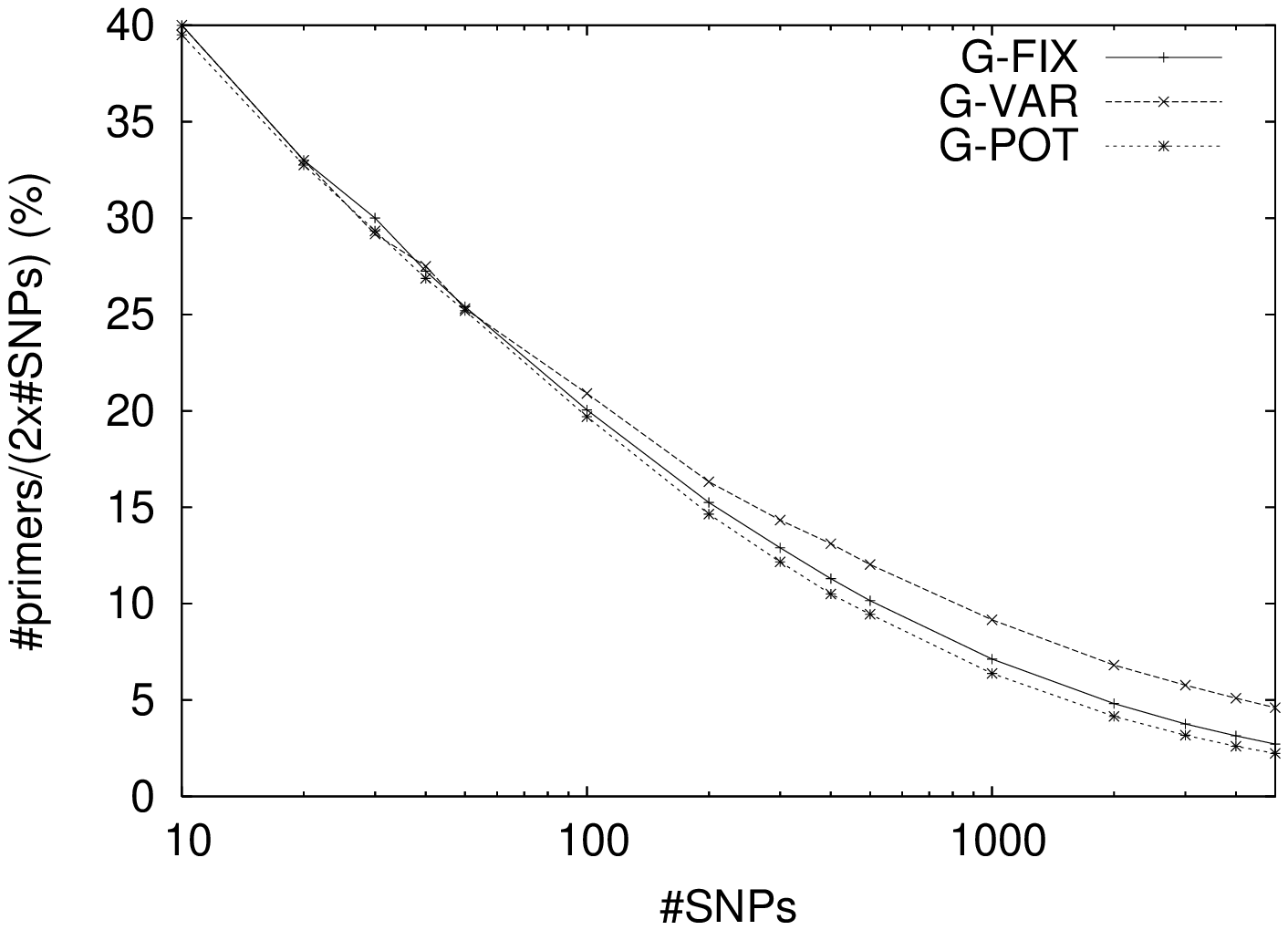,width=\linewidth}\\
(a)
    \end{minipage} 
    \hspace{0.0cm}  
   \begin{minipage}[b]{0.46\linewidth}
      \centering \psfig{figure=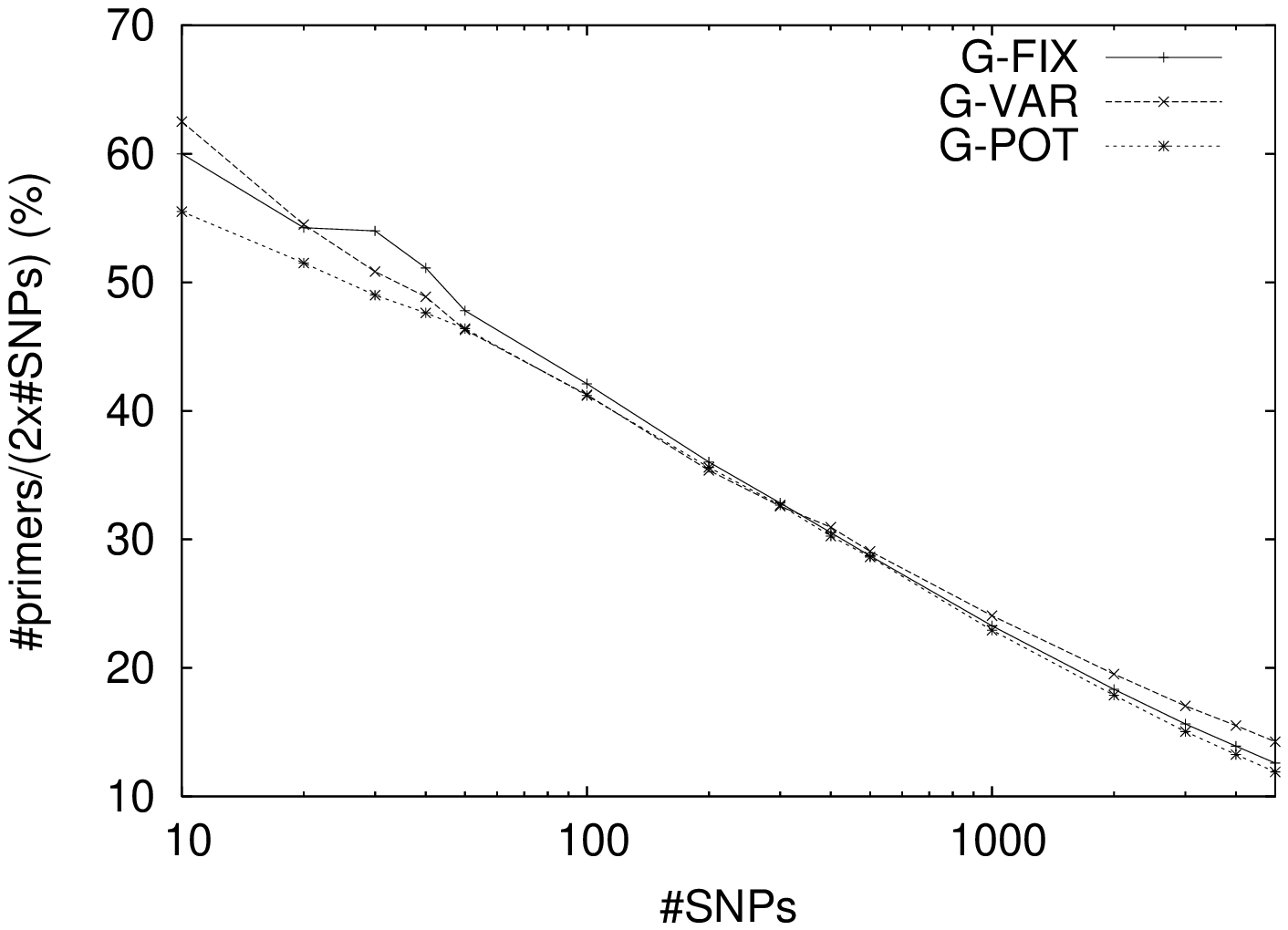,width=\linewidth}\\
(b)
    \end{minipage} \\
    \begin{minipage}[b]{0.46\linewidth}
      \centering \psfig{figure=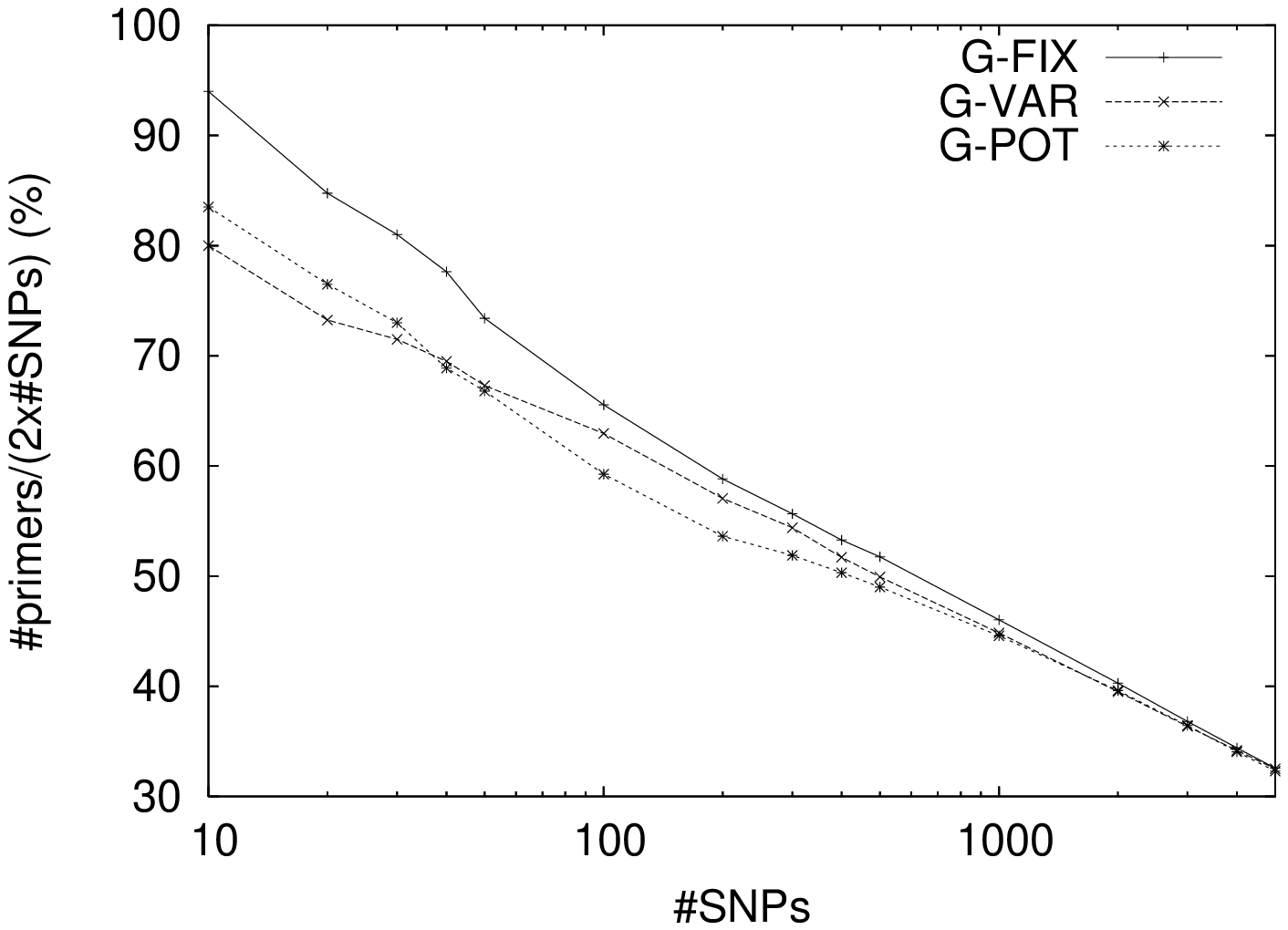,width=\linewidth}\\
(c)
    \end{minipage} 
    \hspace{0.0cm}  
   \begin{minipage}[b]{0.46\linewidth}
      \centering \psfig{figure=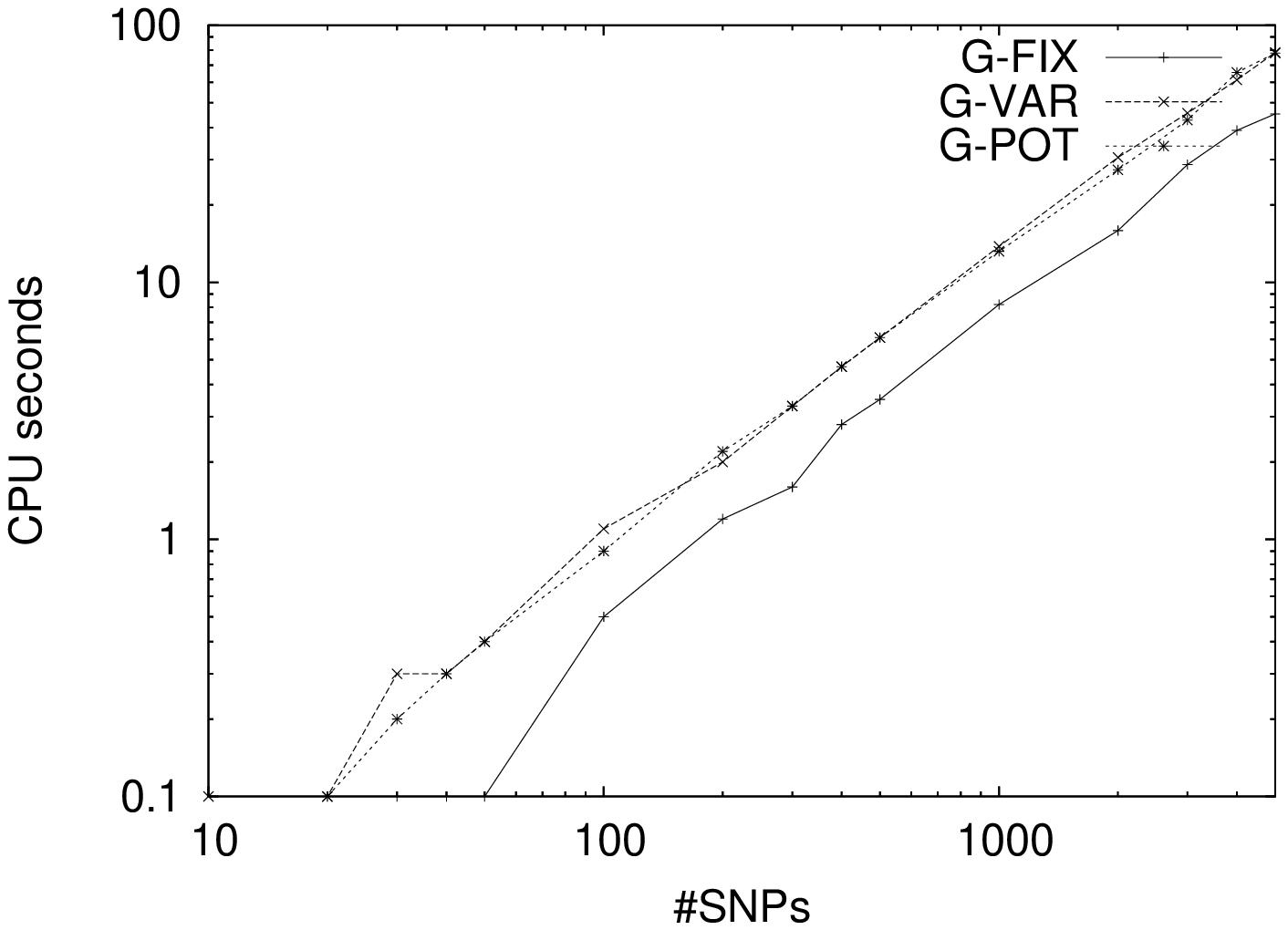,width=\linewidth}\\
(d)
    \end{minipage} 
\caption{(a)--(c) Performance of the compared algorithms,
measured by relative improvement over the trivial solution          
of using two primers per SNP for $k=8, 10, 12$, 
$L=1000$, and up to 5000 SNPs.
(d) Runtime of the compared algorithms for $l=10$,
$L=1000$, and up to 5000 SNPs.
Each number represents the average over 10 testcases of 
the respective size.} 
 \label{primer8-12}
\end{center}
\end{figure}

To further characterize the performance of compared algorithms,
in Figure \ref{primer8-12}(a-c) we plot the
average solution quality of the three greedy algorithms versus
the number of target SNPs (on a log scale)
for randomly generated testcases.
MIPS was not included in this comparison due to its
prohibitive running time.
In order to facilitate comparisons across instance sizes,        
the size of the primer cover is normalized by
the double of the number of SNPs, which is the size of           
the trivial cover obtained by
using two distinct primers to amplify each SNP.                 
Although the improvement is highly dependent
on primer length and number of SNPs,
G-POT is still consistently outperforming
the G-FIX algorithm of\cite{PearsonRWZ96},
and, with few exceptions, its G-VAR modification.

Figure \ref{primer8-12}(d) gives the log-log plot of the average CPU
running time (in seconds) versus the number of pairs of sequences
for primers of size  10 and randomly generated pairs of sequences.
All experiments were run on a PowerEdge 2600 Linux
server with 4 Gb of RAM and dual 2.8 GHz Intel Xeon CPUs -- only one of
which is used by our sequential algorithms -- using the same compiler
optimization options.
The runtime of all three greedy algorithms grows linearly with the number of
SNPs, with G-VAR and G-POT incurring only a small factor penalty in runtime
compared to G-FIX.  This suggests that a robust practical heuristic is
to run all three algorithms and return the best of the three solutions found.

\section{Open Problems}
\label{conclusions.sec}

While the logarithmic approximation factor achieved 
by our greedy algorithm for PCR primer set selection 
with an amplification length constraint of $L$ is 
optimal within an additive factor of $O(\ln{L})$, 
the gap between the $O(\ln{n})$ inapproximability 
bound established in \cite{FernandesS02} and the 
approximation factor of  $O(L\ln{n})$ that we obtain 
for PCR primer set selection
with uniqueness constraints is less satisfactory. 
Closing this gap, either directly or via improved approximations 
for the minimum multi-colored subgraph problem, is an interesting 
open problem.

{\small
\bibliography{primer,longnames}
\bibliographystyle{plain}
}
\end{document}